\documentclass[ reprint, amsmath,amssymb, aps,pra, twocolumn
]{revtex4}

\usepackage{graphicx}% Include figure files
\usepackage{dcolumn}% Align table columns on decimal point
\usepackage{bm}% bold math
\usepackage{amssymb}
\usepackage{mathrsfs}
\usepackage[notransparent]{svg}
\usepackage{svg}
\usepackage{todonotes}
\usepackage{listings}
\usepackage{xcolor}
\usepackage{soul}
\newcommand{\ket}[1]{|#1\rangle}

\newcommand{\abs}[1]{ \left\lvert#1\right\rvert}
\newcommand{\AlO}{\ensuremath{\mathrm{Al}_2\mathrm{O}_3}}
\newcommand{\SiO}{\ensuremath{\mathrm{Si}\mathrm{O}_2}}
\newcommand{\HfO}{\ensuremath{\mathrm{Hf}\mathrm{O}_2}}

\begin{document}

\preprint{APS/123-QED}

\title{Atomic fluorescence collection into planar photonic devices}% Force line breaks with \\

\author{Orion Smedley}
\thanks{These two authors contributed equally}
\author{Vighnesh Natarajan}
\thanks{These two authors contributed equally}
\author{Oscar Jaramillo}
\author{Hamim Mahmud Rivy}
\author{Karan K. Mehta}
 \email{karanmehta@cornell.edu}
\affiliation{%
School of Electrical and Computer Engineering, Cornell University
}%

\date{\today}% It is always \today, today,
             %  but any date may be explicitly specified

\begin{abstract}

%Fluorescence collection from individual dipoles enables state detection and remote entanglement generation, both key to many quantum platforms. To this end, planar photonics promise robust scalable collection, and have already demonstrated the reciprocal process of ion addressing. In this reverse process, we show that a single field component at a single point of a structure's emission quantifies the structure's collection efficiency. This relation connects efficiency to the spot size of gaussian beams, which in the tight focus regime, perform and look similar to optimal mode matched beams. Rather, the losses come, at present, from photonic nonidealities. In the grating we simulate and fabricate, our 0.25 \% efficiency is 8x under the gaussian limit for our $1.9 \lambda$ spot. We lose $\sim 2\times$ each to the second order beam, downward radiation, and side lobes. We point out methods for improvement, as well as a simple scheme enabled by integrated collection for polarization-based remote entanglement generation.

Fluorescence collection from individual emitters plays a key role in state detection and remote entanglement generation, fundamental functionalities in many quantum platforms. Planar photonics have been demonstrated for robust and scalable addressing of trapped-ion systems, motivating consideration of similar elements for the complementary challenge of photon collection. Here, using an argument from the reciprocity principle, we show that far-field photon collection efficiency can be simply expressed in terms of the fields associated with the collection optic at the emitter position alone. We calculate collection efficiencies into ideal paraxial and fully vectorial focused Gaussian modes parameterized in terms of focal waist, and further quantify the modest enhancements possible with more general beam profiles, establishing design requirements for efficient collection. Towards practical implementation, we design, fabricate, and characterize two diffractive collection elements operating at $\lambda=397$ nm; a forward emitting design is predicted to offer 0.25\% collection efficiency into a single waveguide mode, while a more efficient reverse-emitting design offers $1.14\%$ collection efficiency, albeit with more demanding fabrication requirements. Close agreement between simulated and measured emission for both designs indicates practicality of these collection efficiencies, and we indicate avenues to improved devices approaching the limits predicted for ideal beams. We point out a particularly simple integrated waveguide configuration for polarization-based remote entanglement generation enabled by integrated collection.
\end{abstract}  

%\keywords{Suggested keywords}%Use showkeys class option if keyword
                              %display desired
\maketitle

%\tableofcontents

In the context of trapped-ion quantum information processing \cite{bruzewicz2019trapped}, collection of light scattered from atomic ions provides a mechanism for state measurement \cite{dehmelt1982monoion, bergquist1986observation, nagourney1986shelved, sauter1986observation, harty2014high}, as well photonic entanglement of distant qubits \cite{simon2003robust, moehring2007entanglement, stephenson2020high, krutyanskiy2023entanglement, o2024fast, saha2024high}. This functionality is typically implemented by means of a high numerical aperture (NA) objective collecting far-field radiation onto a detector. Pursuit of large-scale trap arrays, as well as compact and robust implementations of full experimental systems \cite{moses2023race}, has motivated recent work towards integration of detector devices within ion trap chips for direct fluorescence capture \cite{todaro2021state, setzer2021fluorescence, reens2022high}. 

For the complementary challenge of optical delivery to ions, waveguides and beam-forming gratings integrated withing ion trap chips \cite{mehta2016integrated, mehta2020integrated, niffenegger2020integrated, ivory2021integrated} offer advantages over conventional free-space approaches in beam-pointing and phase stability \cite{mehta2020integrated,niffenegger2020integrated, vasquez2023control}, along with promise for scalability \cite{mordini2024multi, kwon2024multi}. Via reciprocity \cite{snyder1983optical, haus1984waves, novotny2012principles}, the same structures can be utilized for collection of light emitted from ions into the waveguides. 

Integrated collection into waveguide devices may offer significant benefits in key aspects of scalable, integrated atomic systems. This includes for qubit state readout, and as has been recently discussed \cite{knollmann2024integrated}, remote entanglement generation. Potential advantages include parallelizability, background signal rejection via spatial mode filtering, and ability to locate integrated detectors at regions remote from fluorescing atoms, in principle allowing flexible electromagnetic shielding of the detectors from ions and trap electrodes, as well as decoupling detector area from solid angle subtended by the collection optics \cite{ferrari2018waveguide}. Furthermore, and as discussed below, photon collection into single-mode photonics offers routes to particularly simple, robust implementations of single-photon interference required for entanglement generation between spatially separated ions. 

Resonant cavities offer a route to order unity collection efficiency at rates beyond that set by free-space spontaneous emission \cite{keller2004continuous, wilk2007single, CollectIonCavityMonroe2012, thompson2013coupling,Kim2011CollectionFiber, schupp2021interface, krutyanskiy2023entanglement}; the associated technical challenges though, particularly for ions, motivate an understanding of limits to free-space collection more readily implemented in scalable platforms. Far-field collection efficiency in integrated settings is subject to the same  NA limits as with bulk optics. Only preliminary estimates for collection efficiencies into planar photonic structures have been presented in the literature to date, with limited work towards optimal designs for integrated collection optics. 

To precisely describe collection efficiencies achievable and establish efficient metrics for optimization, here we show that a simple argument from reciprocity allows expression of the polarization-dependent collection efficiency in terms of the electric field that would be radiated by the collection optic projected onto the radiating dipole, enabling straightforward quantification and optimization of photon collection with almost no approximations. We analyze practically achievable values for idealized collection elements designed to couple to Gaussian beams, extending beyond the paraxial approximation as required for the high-NA focuses required for efficient collection. We present preliminary design and characterization of planar single-mode photonic collection optics, demonstrating submicron waist focusing at $\lambda=397$ nm in agreement with designed performance, and with a predicted total 0.25\% collection efficiency. An improved device with suppressed spurious sidelobe scatter is predicted to offer 1.14\% collection efficiency. These considerations indicate the required field concentration is achievable in current designs at the blue/UV wavelengths of interest for most species \cite{knollmann2024integrated}, and we discuss routes to more complete optimization to approach the limits predicted. We also point out a significant simplification possible for polarization-based remote entanglement generation using integrated collection into single-mode photonics and single-photon interference \cite{stephenson2019entanglement}. Our work establishes a simple foundation on which to analyze/design collection elements, and lays a basis for more sophisticated optimal design. 

\begin{figure}[]
    \centering
    \includegraphics[width=\columnwidth]{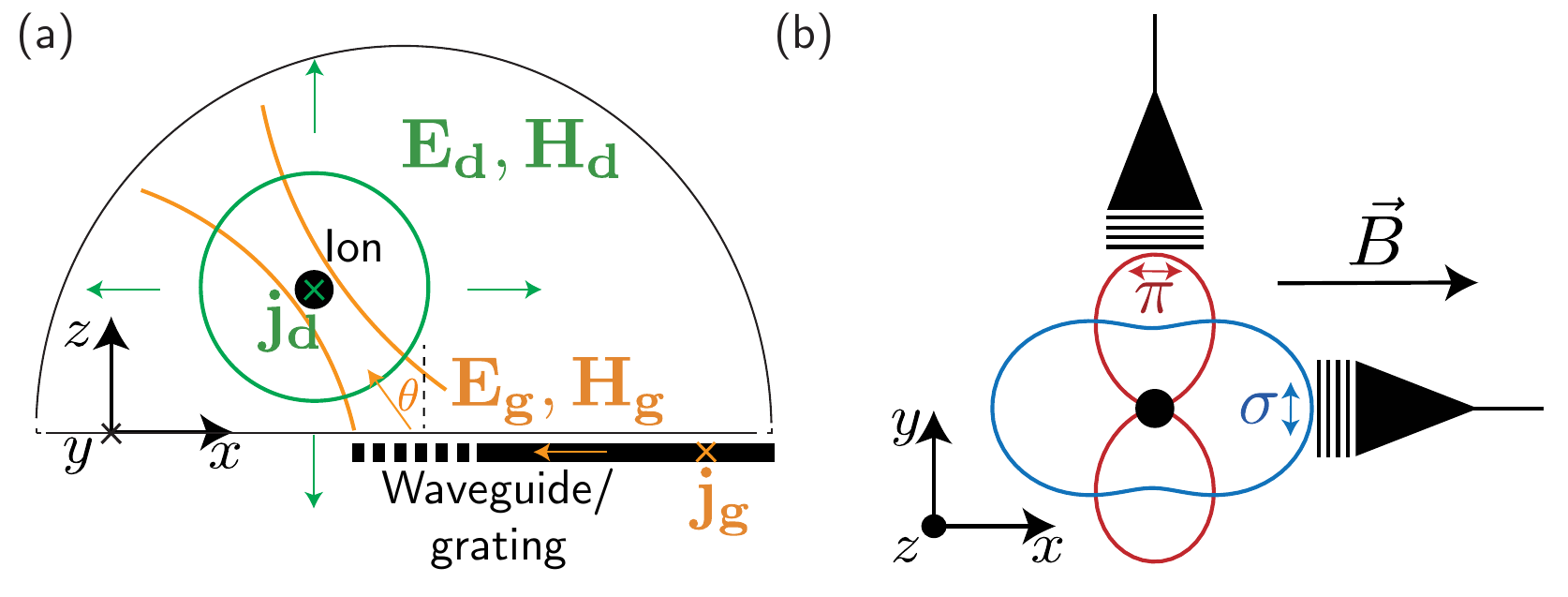}
    \caption{(a) Schematic depicting point dipole source $\mathbf{j}_\mathrm{d}$ corresponding to fluorescing ion and associated fields $\mathbf E_\mathrm{d}$ and $\mathbf H_\mathrm{d}$, along with fields radiated by the collection waveguide and grating as if sourced by an effective current source $\mathbf j_\mathrm{g}$ launching amplitude in the single waveguide mode sourcing the grating. Green/orange arrows (lines) represent power flows (intensity countours) associated with the fields labeled.  The line at the origin defined as $z=z_g$ represents a choice of overlap integral (Eq.~\ref{eq:overlapIntegral}) evaluation plane near the grating.  (b) Orthogonal polarization states radiated by an atom, defined with respect to an external, quantizing B-field, can be collected into identical modes of separate waveguides. Black features represent waveguides and gratings in the $x-y$ plane; double-sided arrows represent the dominant E-field polarization for either quasi-TE waveguide mode. For the orientation depicted,  $\pi$ and $\sigma$-polarized light (with radiation patterns denoted by red and blue solid lines) is coupled into quasi-TE modes of separate waveguide channels.}
    \label{fig:schematic}
\end{figure}

Collection efficiency into a single mode can be expressed in terms of the overlap between the emission pattern of a classical point dipole \cite{jackson1999classical} with that of a beam that would be emitted by a ``collection grating" were it to be illuminated through the single mode into which it couples (Fig.~\ref{fig:schematic}). That is, we have complex fields $\mathbf{E_\mathrm{d}, H_\mathrm{d}}$ associated with the radiation from the point dipole with dipole moment $p_0 \mathbf{\hat p}$ (magnitude $p_0$, unit vector $\mathbf{\hat p}$), and $\mathbf{E_\mathrm{g}, H_\mathrm{g}}$ associated with the field emitted by a grating, both solutions appropriately normalized to unit power (i.e. with units of V/m$\sqrt{\mathrm W}$ and A/m$\sqrt{\mathrm W}$). We work with the convention that the physical field is given by the real part, e.g. for the dipole radiation, by $\mathrm{Re}(\mathbf E_d)$. 
We consider $p_0$ and all fields to be oscillating at a single frequency $\omega_0$ corresponding to the atomic transition's resonance; the single-frequency approximation is valid as long as the dipole radiation profile varies negligibly over the natural linewidth $\Gamma$ of the transition, a valid approximation given that typically utilized atomic transitions have $\Gamma/\omega_0 \ll 10^{-5}$. 

The power coupling $\eta$ between these fields, i.e. the fraction of power radiated by the dipole collected into the waveguide mode, can be written as an overlap integral \cite{snyder1983optical} between the power-normalized dipole and grating fields: 
\begin{eqnarray}
\eta &=& \abs{ \int_{z=z_g} \frac14 \left( \mathbf{E_\mathrm{d}} \times \mathbf{H_\mathrm{g}^*} + \mathbf{E_\mathrm{g}^*} \times \mathbf{H_\mathrm{d}} \right) \cdot d\mathbf a }^2 . \label{eq:overlapIntegral} %\\
\end{eqnarray}
For simplicity, we choose  $z=z_g$, the plane just above the collection grating (Fig.~\ref{fig:schematic}a). However, the reasoning below shows $\eta$ is independent of this choice.

A simpler, equivalent expression for $\eta$ derives from Lorentz reciprocity. Considering a (power normalized) source current associated with the emitting atom $\mathbf{j_\mathrm{d}}$ sourcing $\mathbf{E_\mathrm{d}}$ and $\mathbf{H_\mathrm{d}}$, and $\mathbf{j_\mathrm{g}}$ the effective source for the grating field $\mathbf{E_\mathrm{g}}$ and $\mathbf{H_\mathrm{g}}$, the fundamental reciprocity relation tells us that \cite{novotny2012principles}
\begin{equation}
\nabla \cdot  \left( \mathbf{E_\mathrm{d}} \times \mathbf{H_\mathrm{g}^*} + \mathbf{E_\mathrm{g}^*} \times \mathbf{H_\mathrm{d}} \right) = -\mathbf{j_\mathrm{d}}\cdot\mathbf{E_\mathrm{g}^*} - \mathbf{j_\mathrm{g}^*} \cdot \mathbf{E_\mathrm{d}}. 
\end{equation}
We consider a radiating point dipole at $\mathbf r_0$ with polarization vector $\mathbf{ p}=p_0\mathbf{\hat p}$ so that $\mathbf{j_\mathrm{d}} = -i \omega_0 p_0 \mathbf{p} \delta(\mathbf r-\mathbf{r_0})$, and $\mathbf E_\mathrm g$ the field of the beam emitted by the grating coupler with effective source lying below $z=z_g$. We can integrate over the infinite half-volume above $z=z_g$ and use the fact that in the far-field (on the hemisphere at $r\rightarrow \infty$, depicted in Fig.~\ref{fig:schematic}a), where the propagating fields are transverse, the quantity within the divergence above vanishes at all points \cite{pozar2011microwave} to find

\begin{equation}
\int_{z=z_g}\left( \mathbf{E_\mathrm{d}} \times \mathbf{H_\mathrm{g}^*} + \mathbf{E_\mathrm{g}^*} \times \mathbf{H_\mathrm{d}}\right) \cdot d\mathbf a = i \omega_0 p_0 \mathbf{\hat p} \cdot \mathbf E_\mathrm{g}^*(\mathbf r_0). 
\label{eq:FOM}
\end{equation}
Substituting into the overlap integral (Eq. \ref{eq:overlapIntegral}), the coupling efficiency 
\begin{equation}
\eta = \frac{1}{16}\omega_0^2 p_0^2 \abs{ \mathbf{\hat p} \cdot \mathbf E_\mathrm{g}^*(\mathbf r_0)}^2
\label{eq:field projection}
\end{equation}
is expressed in terms of the  power normalized  $ p_0$ and $\mathbf E_\mathrm{g} $ evaluated only at the ion location.  This form also shows that the overlap integral (Eq.~\ref{eq:overlapIntegral}) is equivalent for any arbitrary plane or curved surface.  
%Because the  overlap integral (Eq.~\ref{eq:overlapIntegral}) agrees, its arbitrary (and possibly curved) integration surface does not impact $\eta$. On that surface, a larger numerical aperture (if well mode matched) increases the overlap integral. Equivalently, it shrinks the focus, which increases the focal field projection. 

Expressing $\eta$ this way establishes a straightforward basis on which to design/optimize collection into single modes of any structure, simply requiring appropriately polarized, tight focuses at the ion location to maximize the (normalized) field projected on the radiating dipole, $\abs{\mathbf{\hat p} \cdot \mathbf E_\mathrm{g}^*(\mathbf r_0)}$. For readout, we simply maximize total efficiency, whereas tailored polarization selectivity quantified by the vector dot product above is key for many schemes for generation of ion-photon and ion-ion entanglement as described above. Additionally, positioning errors' impact on $\eta$ is given by the spatial dependence of the grating field $\mathbf E_\mathrm{g}(\mathbf r)$.

\begin{figure*}[ht!]
    % \centering
    \includegraphics[width= \textwidth]{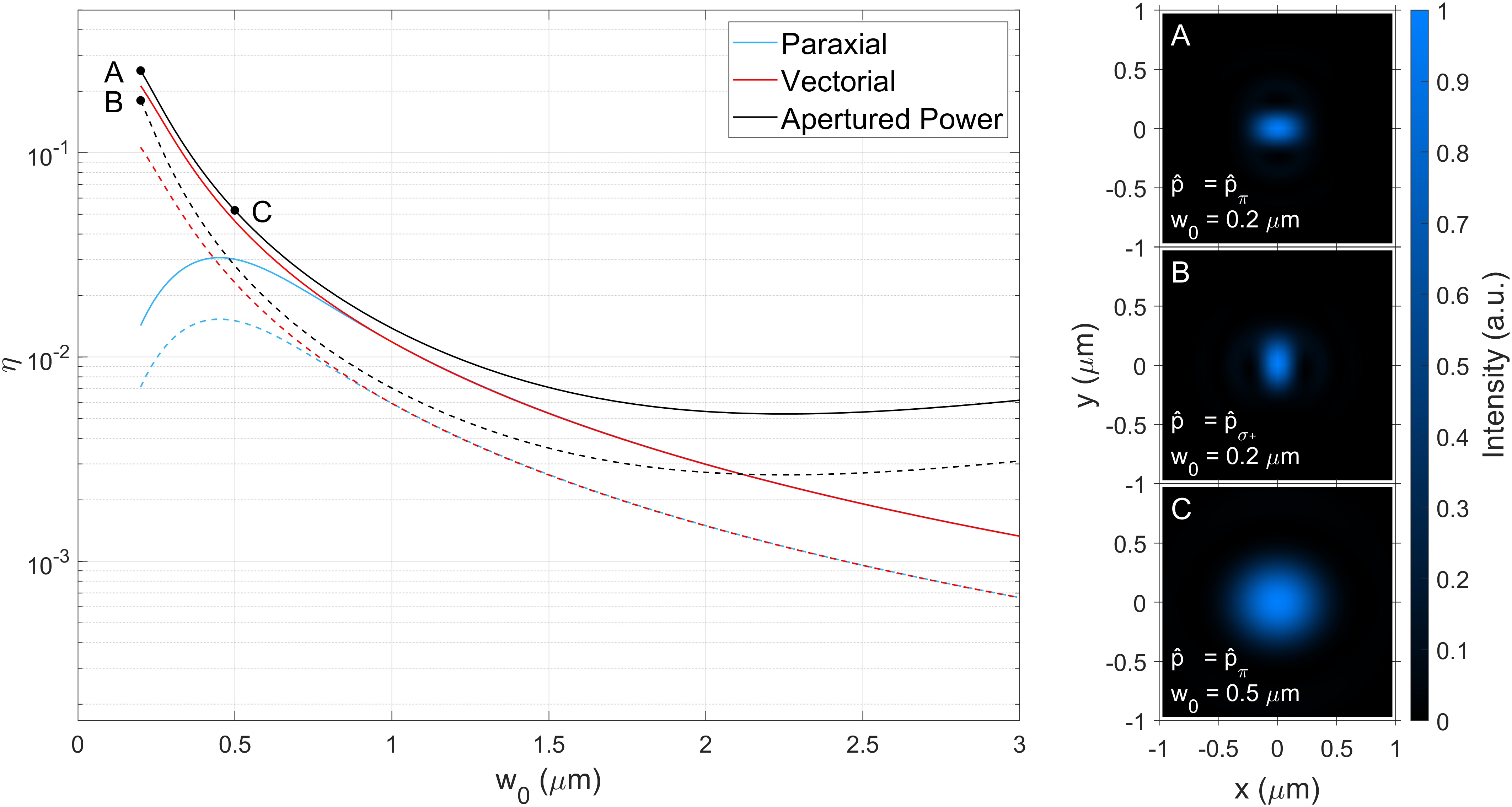}
    \caption{Paraxial and vectorial Gaussian beam coupling to dipole radiation at $\lambda=397$ nm.  $\eta$ is calculated for $\pi$ (solid lines) and $\sigma_\pm$ (dashed lines) radiation (Fig.~\ref{fig:schematic}) coupling to Gaussian beams with dominant linear polarization along $\mathbf{\hat x}$ and $\mathbf{\hat y}$, respectively. Both methods of calculating efficiency $\eta$ (Eq.~\ref{eq:overlapIntegral}, \ref{eq:field projection}) agree for exact vectorial fields (red curve) and are independent of ion height.  Overlap calculated via the paraxial approximation (blue) falls off from the exact result (red) for the small waists relevant for high collection efficiency, shown here for 40 $\mu$m ion height.  Dipole polarization $\sigma_\pm$  (dashed) couples half as well as $\pi$ (solid)  by linearity (see main text) with the exception of the apertured power curves  (black). The optimal beams (black), and the $\pi$ polarized Gaussian (solid red) approach 50\% power coupling (top left) for infinitely tight focuses.  A wavelength-waist Gaussian beam (red) performs 89\% as well as the ideal beam (green) for $\pi$ polarized dipoles, and 80\% optimally for  $\sigma_\pm$.  Corresponding to the apertured power (black dots), the optimal beams profiles in the focal plane are shown at right for the three labeled points A, B, and C in the left panel, and resemble a Gaussian spot both in shape and size, except at the tightest focuses (A,B).}
%    \caption{Paraxial and vectorial Gaussian beam coupling to dipole radiation at $\lambda=397$ nm. Both methods of calculating efficiency $\eta$ (Eq.\ref{eq:overlapIntegral}, \ref{eq:field projection}) agree for exact vectorial fields (red curve). The paraxial approximation retains $\eta$’s qualitative features, when calculated via field projection . These curves are independent of ion height. However, when calculated via the overlap integral, paraxial $\eta$ (blue) falls off from exact $\eta$ (red) for the small waists relevant to high collection efficiency. Solid lines represent $\pi$ radiation coupling into a dominantly $\mathbf{\hat x}$ polarized gaussian, while dashed lines represent $\sigma_\pm$ and $\mathbf{\hat y}$.  $\sigma_\pm$  (dashed) couples half as well as $\pi$ (solid)  by linearity (see main text) with the exception of the apertured power curves  (black). The apertured power (black), and the pi polarized Gaussian (solid red) approach 50\% power coupling (top left) for infinitely tight focuses.  A wavelength-waist Gaussian beam (red) captures 89\% of the apertured power (black) for $\pi$ polarized dipoles, and 80\% for $\sigma_\pm$.  Capturing the full apertured power, the optimal mode profile in the focal plane is shown at right for the three labeled points A, B, and C in the left panel. Optimal profiles resemble a gaussian spot both in shape and size (C),  except at the tightest focuses (A,B).   }
    \label{fig:ReciprocityVerification}
\end{figure*}

As a consistency check between both expressions for $\eta$ and to provide insight into required beam profiles, we numerically calculate $\eta$  according to Eqs.~\ref{eq:overlapIntegral} and~\ref{eq:field projection}. We calculate dipole emission coupling to ideal Gaussian modes represented both paraxially and fully vectorially, as well as the optimal beams for a given numerical aperture constraint. For simplicity, we set the grating beam emission angle to be vertical ($\theta=0$ in Fig.~\ref{fig:schematic}a), that is, propagating along $+z$ with  $\mathbf{E_\mathrm{g}}$ primarily in the $x-y$ plane. 

%This applies to beams rotated by some angle around the quantization axis of the dipole, because the dipole is rotationally symmetric about that axis. The grating plane breaks that symmetry, but the coupling depends only on the focal field. Beams propagating non-orthogonally to the quantization axis will reduce field projection along the dipole vector by the cosine of that angle.

As a particular experimental example we take $\lambda = 397$ nm corresponding to the $S_{1/2} \leftrightarrow P_{1/2}$ transition in $^{40}$Ca$^+$ \cite{bruzewicz2019trapped}, noting however that the scale invariance of Maxwell's equations allow application of these results to other ion species by rescaling for wavelength. 

We orient the magnetic field defining the quantization axis along the $x$-axis  ($\mathbf B \parallel \mathbf{\hat x}$ in Fig.~\ref{fig:schematic}), thereby defining the atomic unit polarization vectors  $\mathbf{\hat p}_\pi = \mathbf{\hat x}$ and $\mathbf{\hat p}_{\sigma_\pm} =(\mathbf{\hat y} \pm i \mathbf{\hat z})/\sqrt{2}$. Next, we scale these polarization vectors $\mathbf{ p} = p_0\mathbf{\hat p}$ to emit unit power with 
\begin{equation}
p_0 \equiv |\mathbf{p}| = \sqrt{\frac{12 \pi}{c^2 Z_0 k^4}}, 
\label{eq: unit power dipole}
\end{equation}
taking the far-field form for the associated fields \cite{jackson1999classical}
\begin{equation}
\begin{aligned}
& \mathbf{H_d}=\frac{c k^2}{4 \pi}(\mathbf{\hat{r}} \times \mathbf{p}) \frac{e^{i k r}}{r} \\
& \mathbf{E_d}=Z_0 \mathbf{H_d} \times \mathbf{\hat{r}}
\end{aligned}
\label{eq: dipole farfield}
\end{equation}
because the distance to a typical grating structure far exceeds the wavelength. Here, $c$ is the vacuum speed of light, $k = 2\pi/\lambda$ is the wavenumber, $Z_0=\sqrt{\mu_0/\epsilon_0}$ is the impedance of free space, and $\mathbf{r} = r \mathbf{\hat{r}}$ is the position vector from the ion location.

For $\mathbf{E_\mathrm{g}}$, we take a Gaussian beam linearly polarized in the $x-y$ plane to maximally couple to $\mathbf{E_\mathrm{d}}$ ($\mathbf{\hat{x}}$ for $\pi$, and $\mathbf{\hat{y}}$ for $\sigma_{\pm}$ emission). We calculate $\mathbf{E_\mathrm{g}}$ both within the paraxial approximation and with a full vectorial treatment.   The paraxial Gaussian propagating along $\mathbf{\hat z}$ takes the textbook form
\begin{align}
E_x(x, y, z) = & \sqrt{\frac{2}{\pi}} \sqrt{2 c \mu_0} \times \frac{\exp\left(-i k z + i \Psi(z)\right)}{w(z)} \nonumber \\
& \times \exp\left(-\frac{x^2 + y^2}{w(z)^2} - i k \frac{x^2 + y^2}{2R(z)}\right),
\label{eq: Paraxial Gaussian}
\end{align}
normalized to unit power in SI units.  The Rayleigh range $z_R$, Gouy phase $\Psi(z)$,  beam waist $w(z) = w_0\sqrt{1+(z/z_R)^2}$, and radius of curvature  $R(z)$ are as defined in \cite{SiegmanLasersBook}. The blue lines in Fig.~\ref{fig:ReciprocityVerification} show the overlap of these paraxial beams evaluated with $\pi$- and $\sigma_\pm$- polarized dipole radiation, for a beam focused at the ion location and for varying focal waist. Note that both $\sigma_\pm$ couple equally to the Gaussian due to its assumed linear polarization, and with half the $\eta$  of $\pi$ radiation since only the $\mathbf{\hat y}$-component is coupled. The increasing $\eta$ with decreasing waist is as expected due to the stronger $\mathbf E_\mathrm g$ at the ion location for tighter focuses. The apparent drop-off for wavelength-scale waists, however, arises due to a focal shift inherent to the breakdown of the paraxial approximation; the paraxial fields at the overlap plane correspond, if propagated exactly, to a focus before the ion. 

To obtain the exact $\eta$ values and verify the  overlap integral for tight focuses of practical interest, we go beyond the paraxial approximation. To do so, we choose the particular solution to Maxwell's equations most similar to the paraxial field profile at the focal plane of the ion by finding its angular-spectrum decomposition \cite{NanoOpticsBook, wolf1959electromagnetic}, then removing longitudinal polarizations from each plane-wave component.  The analytic expression for this Fourier field (equation \ref{eq: Gaussian Angular Spectrum}) is derived in appendix \ref{sec: FFT propagation}.
Phase evolution and subsequent inverse Fourier transform then gives the field profile at any other plane (appendix \ref{sec: FFT propagation}).

With this fully vectorial focused beam, we calculate $\eta$ via the overlap integral (Eq.~\ref{eq:overlapIntegral}) and field projection (Eq.~\ref{eq:field projection}). Both calculations produce the same red lines in Fig.~\ref{fig:ReciprocityVerification} , and confirm $\eta$'s scaling with $w_0$ beyond the paraxial approximation's validity. These results indicate wavelength-scale Gaussian waists provide $\eta$ values of multiple percent. We note that calculating $\eta$ via field projection (Eq. \ref{eq:field projection}) with the paraxial focal fields avoids the paraxial focal shift issue, and gives results within 40\% of the exact values down to the tightest waists shown in Fig.~\ref{fig:ReciprocityVerification}. 

%Second, the Gaussian beam couples twice as much to $\pi$ than $\sigma_\pm$ because the overlap integral (eq. \ref{eq:overlapIntegral}) is linear in $\mathbf{E_\mathrm{d}}$ .  This linearity means a $\sigma$ polarized field  $\mathbf{E_\mathrm{d}}$  is the sum of fields from two orthogonally polarized dipoles. That is $\mathbf{\hat p}_{\sigma_-} = ( \mathbf{\hat p_1} +    i      \mathbf{\hat p_1})/\sqrt{2}  $ , where  $ \mathbf{\hat p_1} = \hat y$ and $\mathbf{\hat p}_2 = \hat z$, each radiating half the power. Only one of the dipoles, the y-polarized one, overlaps with a Gaussian beam,  so the coupling is half. The other dipole, oscillating in the z direction contributes nothing because of symmetry; it cannot distinguish between a positive or negative phase in the y-polarized Gaussian beam.  This is more easily seen through the field projection (eq. \ref{eq:field projection} ) where the Gaussian beam electric field projects out only the x or y components of the dipole polarization.  Note, this factor of two constraint does not apply for optimal coupling (black) discussed next, because linearity does not apply; the optimal field $\mathbf{E_\mathrm{g}}$, in addition to $\mathbf{E_\mathrm{d}}$, both depend on $\hat{\mathbf{p}}$.

We also explore the theoretical maximum performance of an optimal collection optic, which would of course mode-match the dipole emission over its aperture ($\mathbf{E}_\mathrm g = \mathbf{E}_\mathrm d$), thereby coupling the full incident power. For our Gaussian beams we define a relevant aperture as containing 99\% of the power. For the field of Eq.~\ref{eq: Paraxial Gaussian} this corresponds to a radius $1.5 w(z_g)$, where $w(z_g)$ is the beam waist in the grating plane at $z_g$, which of course grows large for tight waists in the focal plane (consistent with numerical aperture). The black lines in Fig.~\ref{fig:ReciprocityVerification} show the total incident radiation over this aperture. These curves demonstrate that a linearly-polarized Gaussian beam focused to $w_0 = \lambda$ performs 89\% as well as the ideal beam given the aperture (black) for $\pi$ polarized dipoles, and 80\% optimally for  $\sigma_\pm$. Note that the increase in apertured power visible in Fig.~\ref{fig:ReciprocityVerification} for large $w_0$ is due to the fact that, in the limit of large focal waists, $w(z_g) \sim w_0$ and hence the apertured power increases with $w_0$ despite poor mode matching between $\mathbf{E}_\mathrm{g}$ and $\mathbf E_\mathrm{d}$.

We also compute the optimal field profiles at the ion plane corresponding to the ideal collection optic. Because $\eta$ is given by the field projection (Eq. \ref{eq:field projection}), this corresponds simply to  maximizing the field at the ion location for a constrained aperture. 
%This differs from optimal focusing literature \cite{ShimMaximalFocusing} in that we have no constraint for the small spatial extent which is required for sub diffraction laser writing or imaging. Rather, we must optimize the field at a single point,  tolerating arbitrary side lobes and non circular spots. 
To do this we simply set $\mathbf{E_\mathrm{g}}=\mathbf{E_\mathrm{d}}$ within the aperture, and 0 elsewhere. We then renormalize the field to unit power, apply time reversal, and propagate the field back to the ion location (appendix \ref{sec: FFT propagation}).  We confirm that this is the ideal field profile by verifying that the field projection (Eq. \ref{eq:field projection}) at the ion location matches the apertured power (labeled points in Fig. \ref{fig:ReciprocityVerification}). 

The optimal intensity profiles are shown in the right panels in Fig. \ref{fig:ReciprocityVerification}. The beam appears qualitatively similar to a Gaussian beam for $w_0 \gtrsim \lambda$, consistent with the Gaussian beam $\eta$ values (red) closely tracking the total apertured power (black). That continues up to impractically wide Gaussian beams ($\approx 2$ $\mu$m), because after that point the aperture on the grating plane begins to grow again. %Then the apertured power becomes the same as the tight focuses. 
These visualizations may offer some intuition for strictly optimal field profile design for collection, though again we note our analysis shows for wavelength-scale beam waists the expected gain beyond focused Gaussians is modest.

\begin{figure*}[ht!]
    \includegraphics[width=1.\textwidth]{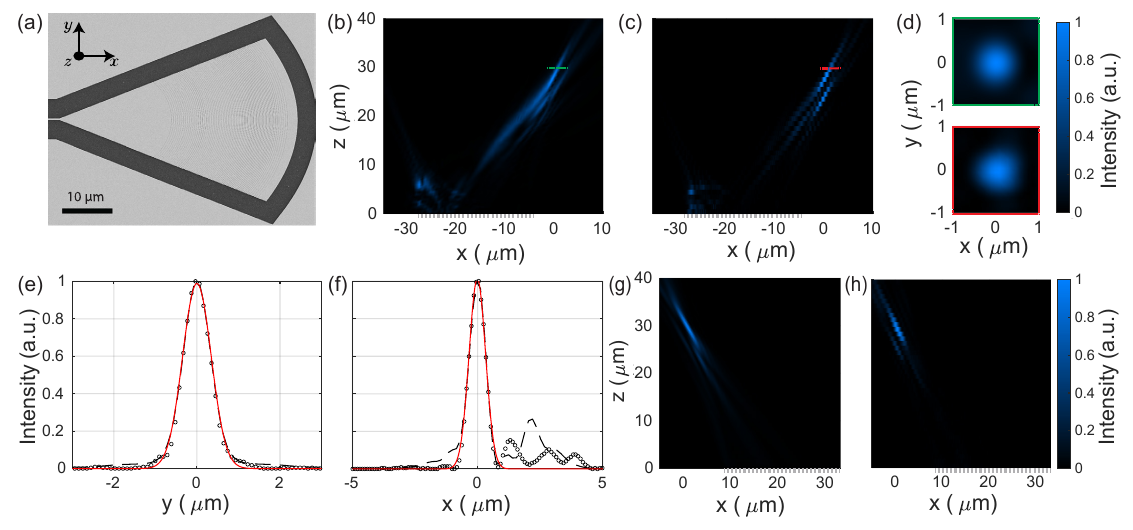}
  \caption{ (a) SEM image of the fabricated grating coupler designed for 30$^\circ$ forward emission with submicron waist at 30 $\mu$m height above the chip surface. Cross sections of simulated (b) and measured (c) beam profiles at $\lambda=397$~nm in the $x-z$ plane, showing multiple diffraction orders. (d) Cross section of the simulated (top) and the measured (bottom) beam profiles, respectively in the $x-y$ plane at the focus height at $z=30$ $\mu$m (within 1 $\mu$m  height measurement uncertainty). (e) and (f) Beam profiles along transverse ($y$) and longitudinal ($x$) cuts, respectively, at the focus. Points are from the measurement, dashed line indicates FDTD simulation result. Red lines are Gaussian fit to the center lobe indicating $w_y = 0.67$ and $w_x = 0.78$ $\mu$m. Simulated (g)  and measured (h) beam profile for single-order reverse-emitting grating. The spatial extent of the grating is indicated by the dashed gray line along the $x$ axis of (b), (c), (g), and (h). Measured data in (c) and (h) are obtained with 1 $\mu$m resolution along $z$. }
  \label{fig:design}
\end{figure*}

% \begin{figure*}[t]
%     \centering
%     \includegraphics[width=1.1\textwidth]{GratingSimPics/IonCollectionFig4_5_4_template.pdf}
%     \caption{ (a) SEM image of the fabricated grating coupler designed for 30$^\circ$ forward emission with submicron waist at 30 $\mu$m height above the chip surface. (b) and (c) Cross sections of simulated and measured beam profiles at $\lambda=397$~nm, respectively in the $x-z$ plane, showing multiple diffraction orders. Insets show intensities at the $z=30$ $\mu$m plane (within 1 $\mu$m  height measurement uncertainty). (d) and (e) Cross section of the simulated and the measured beam profiles, respectively in the $x-y$ plane at the focus height. (f) and (g) Beam profiles along the transverse ($y$) and longitudinal ($x$) cuts, respectively at the focus. Points are from the measurement, dashed line indicates FDTD simulation result. Red lines are Gaussian fit to the center lobe indicating $w_y = 0.67$ and $w_x = 0.78$ $\mu$m. (h) Simulated beam profile for reverse-emitting grating with improved efficiency. }
%     \label{fig:design}
% \end{figure*}

We frame the above discussion by spot size, in order to work independently from the position of the aperture. However, there are some details that do depend on the location $z_g$. While less general, the following still provides an approximate rule of thumb. Given a circular aperture $40$ $\mu$m below the ion and of practical size ($\gg \lambda$), the optimal fill factor for the Gaussian beam is to have a radius that is 88.1\% of the aperture size. This size trades off the power clipped at the aperture with the improved mode matching to maximize the field projection at the focus. We compute this by normalizing the beam power, before clipping the field to 0 outside the aperture. %Note this linear relation between waist and aperture bottoms out at very small radii on the grating plane, because decreasing the beam width further would cause the beam to diverge when exiting the grating.

To compare performance achievable with planar optic collection elements with that predicted for an ideal target Gaussian beam, we design, fabricate and test a focusing collection grating intended for sub-micron spots at $\lambda=397$ nm and a focal height of 30 $\mu$m, using the approach to grating apodization, chirp, and line curvature described in \cite{beck2024grating}. A single 80 nm-thick layer of $(\HfO)_x(\AlO)_{1-x}$ composite \cite{HfO2Jaramillo23}  with $n=1.967$ at $\lambda=397$ nm is used as the waveguide core. In the full stackup this core sits on 3~$\mu$m of thermal $\SiO{}$ on Si substrates, with $1.5$~$\mu$m of top SiO$_2$ cladding. 

A first design targeted a 30$^\circ$ average forward emission angle, to produce constructive interference upon reflection from the Si substrate \cite{mehta2017precise} for the thermal \SiO{} thickness employed here, and a focal waist along both dimensions of $0.5$ $\mu$m. A minimum feature size of 75 nm was used for the design. The average periodicity is ${\sim}350$ nm, with perturbation duty-cycles in the range of 0.2-0.4. 

Fig.~\ref{fig:design}(a) shows an SEM image of the fabricated device. Full 3D finite-difference-time-domain (FDTD) simulation of the structure (Fig.~\ref{fig:design}b and d, upper panel) indicates emission into the targeted Gaussian beam at a focus at $z=30$ $\mu$m with a simulated waist of approximately $0.7$ $\mu$m. The fabricated grating was characterized in beam profiling measurements via emission imaging through a $\mathrm{NA}=0.95$ 50$\times$ objective. The good correspondence between experimentally measured emission and fit waists with the FDTD simulation (Fig.~\ref{fig:design}e,f) indicates that submicron waists as required for efficient collection are realizable with the design method used. 

For a Gaussian beam with $w_0=0.7$ $\mu$m, close to that realized by the fabricated emitter,  we expect $\eta\approx2\%$ for $\pi$-polarized emission (Fig.~\ref{fig:ReciprocityVerification}). However, from the normalized full radiated field as simulated, Eq.~\ref{eq:field projection} predicts an $\eta$ of only $0.25\%$. This ${\sim}8\times$ lower efficiency is due both to the grating's limited upwards radiation efficiency of 47\%, and the appreciable power radiated into sidelobes and additional diffraction orders (Fig.~\ref{fig:design}b,c); for this device the main Gaussian accounts for only 28\% of the upwards radiated power and 13\% of the total input power. Radiation efficiency and scattering into sidelobes and higher orders are key limitations to alleviate to enable efficient integrated collection. 

Higher-order emission can be addressed most simply through emission at angles supporting only a single diffracted order \cite{mehta2017precise}. An improved design targeting the same focal waist in design and supporting only a single diffracted order at a $30^\circ$ reverse angle exhibits a FDTD-simulated total upwards efficiency of $56\%$ with significantly suppressed sidelobes (Fig.~\ref{fig:design}g). This results in a much enhanced predicted $\eta = 1.14\%$ collection efficiency, essentially lower than the values in Fig.~\ref{fig:ReciprocityVerification} only due to upwards radiation efficiency. The smaller periodicities required for diffraction at this angle result in smaller minimum feature sizes of 45 nm and somewhat more challenging fabrication, although roughly consistent with capabilities of advanced foundry photonics processes \cite{rakowski202045nm}. The measured beam profile for a device fabricated to this design is shown in Fig.~\ref{fig:design}h. The fit beam waist of the measured grating emission is $w_x = 0.82$~$\mu$m and $w_y = 0.6$~$\mu$m, in agreement with the full FDTD-simulated waists of $w_x = 0.82$~$\mu$m and $w_y = 0.66$~$\mu$m.

These observations indicate the challenges towards reaching the limits predicted in Fig.~\ref{fig:ReciprocityVerification} with practical devices. While upwards radiation efficiency can be significantly higher than 50\% with use of a bottom reflector \cite{mehta2017precise, beck2024grating}, the range of emission angles involved for tight focusing precludes constructive interference over the full grating area using this simple mechanism, limiting upwards radiation efficiencies in the designs presented here. Two-layer gratings may enable efficient upwards-radiation efficiency over the full aperture even for tight focuses \cite{notaros2016ultra, michaels2018inverse, vitali2023high}, further allowing suppressing of higher-order emission for more relaxed minimum feature sizes as compared to single-layer gratings. In addition, we note that the design methodology itself was based on paraxial propagation which breaks down for the targeted focal spots. This resulted in a larger realized waist both in FDTD simulation and measurement than the design target, and we expect a more accurate design process, or direct optimization for $\eta$, will enable tighter focuses. Addressing these limitations will be critical to realizing integrated single-mode collection efficiencies competitive with that of bulk free-space optics. 

\begin{figure}[t]
    \centering
    \includegraphics[width=\columnwidth]{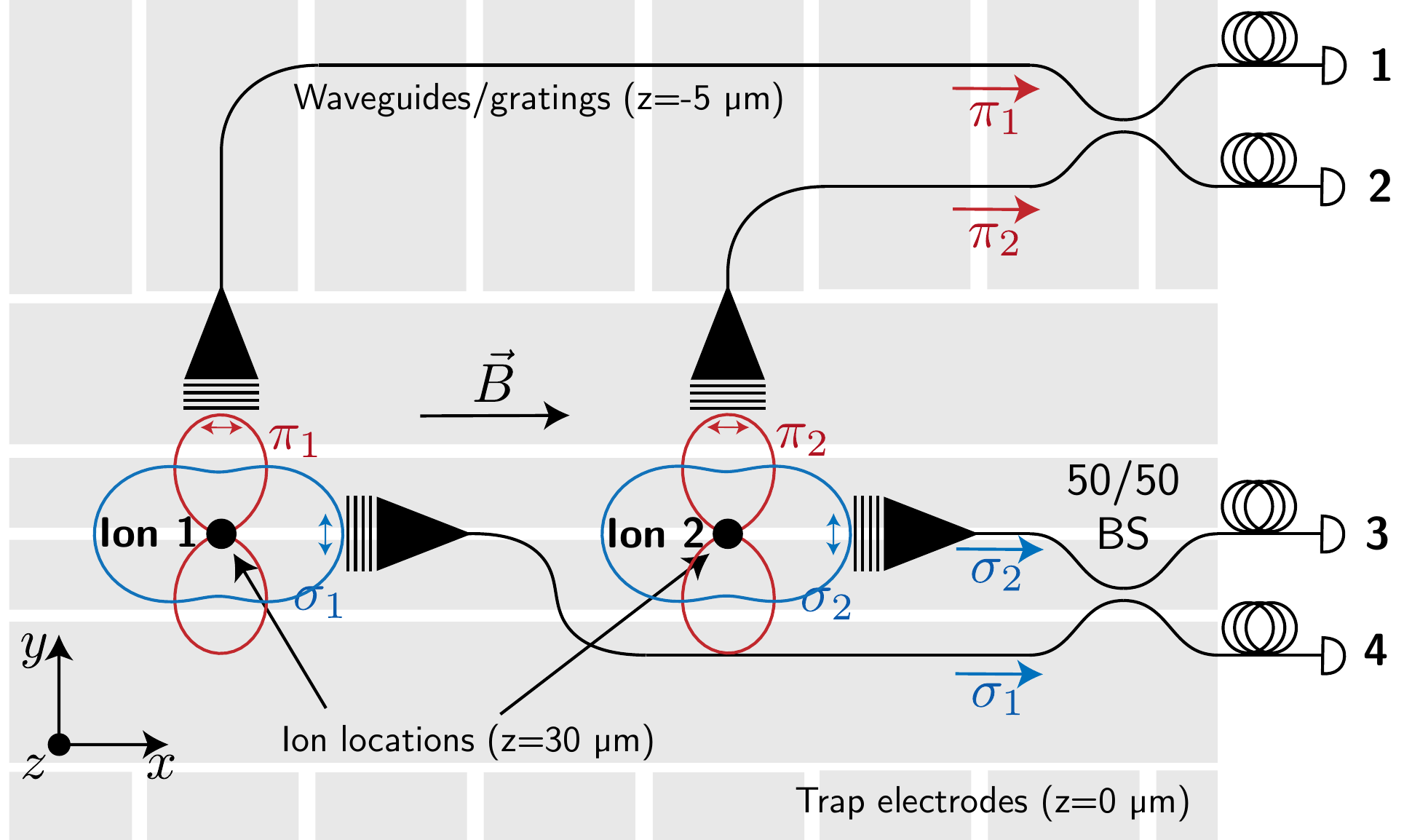}
    \caption{Scheme for integrated polarization-based photon-mediated entanglement generation. Because in the typical Bell-state analyzers used, orthogonal polarizations do not interfere, collection of $\pi$- and $\sigma$-polarized photons into the same modes of separate waveguides, in essentially a dual-rail encoding of the emitted photon state from either atom, allows for the required single-photon interference realized with simply two on-chip 50/50 beamsplitters (BS) acting only on the fundamental quasi-TE waveguide mode. Single-photon detectors 1-4 depicted at the right may be integrated or fiber-coupled off-chip as shown.}
    \label{fig:PMEscheme}
\end{figure}

Finally, we point out that integrated polarization-dependent collection from separate ions into the same fundamental waveguide modes of different waveguides (essentially implementing a ``dual-rail" encoding of the photon as emitted from either atom) as illustrated in Fig.~\ref{fig:PMEscheme} enables a particularly simple implementation of the required single-photon interference for polarization-based remote entanglement generation \cite{stephenson2019entanglement, stephenson2020high, knollmann2024integrated}. Both ions are simultaneously excited and may decay to two possible final states via emission of a $\sigma$- or a $\pi$-polarized photon. Adopting a phase convention with the action of a single beamsplitter  represented by
\begin{equation*}
\begin{pmatrix} b_{1} \\ b_{2}  
    \end{pmatrix} = 
    \begin{pmatrix} 
   -r & it \\
   it & -r 
   \end{pmatrix}
    \begin{pmatrix} 
 a_{1} \\ a_{2} 
    \end{pmatrix}
\end{equation*}
with $a$ and $b$ coefficients labeling inputs and outputs, respectively, and $r=t=1/\sqrt2$ for a 50/50 beamsplitter \cite{haus1984waves}, the two 50/50 beamsplitters clearly implement the transformation
\begin{equation*}
\begin{pmatrix} b_{\pi,1} \\ b_{\sigma, 1} \\ b_{\pi,2} \\ b_{\sigma,_2} 
    \end{pmatrix} = \frac{1}{\sqrt{2}}
    \begin{pmatrix} 
   -1 & 0 & i & 0 \\
   0 & -1 & 0 & i \\
   i & 0 & -1 & 0 \\
   0 & i & 0 & -1 
   \end{pmatrix}
    \begin{pmatrix} 
 a_{\pi,1} \\ a_{\sigma, 1} \\ a_{\pi,2} \\ a_{\sigma,_2} 
    \end{pmatrix}
\end{equation*}
which is exactly that implemented by the 50/50 non-polarizing beamsplitter in the free-space implementation of \cite{stephenson2020high}. With the typical requirement that detection at the beamsplitter outputs encodes no information about the path traveled prior to the beamsplitter, coincident counts on detectors 1 or 2 and 3 or 4 correspond to a photonic state expressed in terms of the modes before the 50/50 beamsplitters of $\left(\ket{1_{\pi_1} 1_{\sigma_2}} \pm \ket{1_{\sigma_1} 1_{\pi_2}}\right)/\sqrt 2$ (with $\ket 1$ representing a single-photon Fock state of the corresponding mode), and a corresponding maximally entangled state of the two emitting atoms for an internal level structure as used in \cite{stephenson2020high}. Since orthogonal polarization states do not interfere in any case, identically polarized photons from the two ions can be separately interfered and detected, and the mode multiplexers and mode-agnostic beamsplitters discussed for waveguide implementations of this scheme in \cite{knollmann2024integrated} can be avoided at the cost of one more splitter acting on the same waveguide modes. This results in a significant simplification of the required photonics, requiring only the collection gratings and $2\times2$ splitters robustly implementable with standard muli-mode-interference devices \cite{soldano1995optical}, acting on a single fundamental waveguide mode for all channels. That $\pi$ and $\sigma$ photons are detected on separate branches automatically ensures that the partial Bell state analyzer is realized. 

Our work points to the requirement for optimizing the field output of a collection device projected along the radiating dipole polarization, to optimize photon collection into a particular mode. This offers a simple figure of merit for optimization, including via inverse design \cite{lalau2013adjoint}, quantification of coupling to undesired modes and polarizations, and of positioning errors on $\eta$. We relate this efficiency to the waists of Gaussian modes, defining spot sizes required to achieve a particular $\eta$. For tight focuses, we show that Gaussian beams perform near (80-90\%) optimally and qualitatively resemble the optimal beams. For practically achievable focuses, we predict $\eta$ on the level of multiple percent. However we note that significant enhancements in entanglement rate are not expected for integrated implementation as compared to high-NA bulk optics in state-of-the-art experiments \cite{stephenson2020high, saha2024high}. In fact our work points to improvements in design required to achieve collection efficiencies into integrated elements competitive with leading bulk optical demonstrations

 Integrated collection as analyzed here with efficient collection optics would enable routing to integrated detectors with high modal selectivity and effective background suppression, as well as the use of compact waveguide-coupled integrated detectors with dimensions independent of the collection aperture. While our discussion has focused on diffractive grating couplers for coupling into SM waveguides, similar functionality may be implemented with devices leveraging metasurface techniques as well \cite{huang2023leaky, zhang2020low, ollanik2024integrated}. For remote entanglement, the parallelizability of integrated collection offers routes to multiplexed generation on multiple ion pairs in general, as well as alternatives to transport-based multiplexing for increased rate given collection efficiency constraints \cite{you2024temporally}. The robust polarization encoding enabled and ability to utilize single modes of separate collection waveguides is expected to translate into significantly higher robustness and perhaps fidelity. And, collection efficiency associated with a given optic can be doubled by coupling into two symmetrically placed elements, and combining their outputs via a 50/50 waveguide splitter. Additionally, collection into field profiles with zero gradient at the ion location, e.g. into standing wave profiles associated with such symmetrically placed emitters, may in the future further enable suppression of collection of sideband scattered photons \cite{mundt2002coupling, vasquez2023control}, which together with the precise polarization and phase control in integrated settings may in the long run assist in optimizing remote entanglement fidelity. 

\vspace{-10pt}
\begin{acknowledgments}
We acknowledge support from an NSF CAREER award (2338897), NSF award no. 2301389, IARPA via ARO through Grant Number W911NF-23-S-0004 under the ELQ program, a Corning Fellowship, and Cornell University. 

We thank Gillenhaal Beck for sharing device design code, Francesco Monticone for helpful discussions, and John Chiaverini and Jonathan Home for comments on the manuscript. 

We note that shortly after posting a preprint version of this manuscript we became aware that a group at Oxford had independently arrived at and very recently put forward a similar observation on interference required for polarization-based remote entanglement generation \cite{ainley2024multipartite}. 
\end{acknowledgments}

\vspace{5pt}
 \textbf{Disclosures: } KKM serves as an advisor to Oxford Ionics (I,E). 

\textbf{Disclosures: } The authors declare no conflicts of interest. 
\section*{Author Contributions}
OS developed and performed the pure beam calculations and analysis. VN designed and simulated the grating devices, and analyzed characterization results with support from OJ and HMR in fabrication and measurement. KKM conceived and supervised the work. OS, VN, and KKM prepared the manuscript with input from all authors. 

%\nocite{*} % Delete this later
\providecommand{\noopsort}[1]{}\providecommand{\singleletter}[1]{#1}%

\appendix

\section{FFT propagation} \label{sec: FFT propagation}

We propagate fully vectorial fields from one plane to another via an angular spectrum decomposition \cite{NanoOpticsBook, wolf1959electromagnetic}. First, we describe the method and how it describes propagation fully obeying Maxwell's equations, and finally we explain its numerical implementation with the fast Fourier transform.

\subsection{Angular spectrum and Maxwell Corrections}

The angular spectrum is the 2D Fourier transform of the field on a plane. That is,

\begin{align}
    \widetilde{\vec{E}}|_{z_{i}} &:= \mathscr{F}_\text{2D}[\vec{E}|_{z_{i}}]; \\ 
    \widetilde{\vec{E}}|_{z_{i}} (k_x,k_y) &= \frac{1}{\sqrt{2\pi}} \int_{x=-\infty}^{\infty} \int_{y=-\infty}^{\infty} \vec{E}(x,y,z_{i}) \notag \\
    &\quad \times \exp\left[-i(k_xx + k_yy)\right] \notag\\
    &\quad \times\exp\left[-i k_z z_{i}\right] \, dx \, dy
    \label{eq: angular spectrum}
\end{align}
formally and in components respectively. Here, the $z_{i}$ is the initial plane that the angular decomposition is performed on, and $k_z=\sqrt{k_0^2 -k_x^2-k_y^2}$. 

Second, we apply this decomposition to compare the paraxial Gaussian to the closest exact solution to Maxwell's equation. The paraxial Gaussian (Eq. \ref{eq: Paraxial Gaussian}), along with most arbitrary fields, do not satisfy Maxwell's equations.  Specifically, their angular spectrums' include longitudinal waves. Subtracting these components from the original Fourier field $\widetilde{E}$ yields the corrected angular spectrum $\widetilde{E}^c$: $$\widetilde{E}^c:=\widetilde{\vec{E}}-(\widetilde{\vec{E}} \cdot \hat{k}) \hat{k}.$$
Next,  we remove the evanescent waves to improve numerical stability. This does not affect the coupling, because these waves do not carry power to the collector in the far field.  Explicitly we remove evanescent waves by
$\widetilde{\vec{E}}^{c e}=\left\{\begin{array}{ll}
    \widetilde{\vec{E}}^c & \text { if } k_{x y}<k_0 \\
    0 & \text { else }
    \end{array}\right\}$
After performing these two corrections, we renormalize the field to unit power.

In order to compute the field along other planes, we compute the inverse Fourier transform of an angular decomposition that has advanced in phase. That is 
\begin{equation}
    \vec{E}|_{z_f}:=\mathscr{F}^{-1}_\text{2D}\left[\widetilde{\vec{E}}^{c e} ~ \exp[ {i k_{z} \cdot \left(z_f-z_i\right)  ]}\right]
    \label{eq:debye propagator}
\end{equation}
where $z_f$ is the final plane that we want the real space vector field. In the general case, the FFT will compute this with complexity $\mathcal{O}(n \log n)$, where $n$ is the number of sample points.  In special cases, this can be done analytically, such as the paraxial Gaussian beam. Starting from its focal field, to preserve the spot size, we analytically solve for $\widetilde{\vec{E}}^{c e}$.  The closed form solution for a primarily $x$-polarized fully vectorial Gaussian is then
\begin{align}
     \widetilde{\vec{E}}_{\text{vG}}(k_x,k_y) |_{z = 0}&= \\
     \frac{w_0 e^{-\frac{1}{4} w_0^2 
     \left(k_x^2+k_y^2\right)}}{8 \sqrt{2} \pi ^{7/2}} 
    &\begin{bmatrix}
        -k_{x}^2 \lambda^2 + 4\pi^2 \\
        -k_{x} k_{y} \lambda^2 \\
        -k_{x} \lambda \sqrt{-(k_{x}^2 + k_{y}^2) \lambda^2 + 4\pi^2}
    \end{bmatrix}
    \label{eq: Gaussian Angular Spectrum}
\end{align}
where $k_x,k_y,$ and $k_z = \sqrt{{(2 \pi / \lambda)}^2 -k_x^2-k_y^2}$ are wave-vector components, $w_0$ is the focal radius, and $\lambda$ is the wavelength of light. This field rigorously satisfies Maxwell's equations as a sum of transverse plane waves, as demonstrated by the dot product $\widetilde{\vec{E}}_{\text{vG}} \bullet \vec{k} =0$. 

\section{Numerical Details}
The overlap integral is numerically calculated along a sample grid. We sample the field along the grating plane over a $200$ $\mu$m $\times$ $200$ $\mu$m domain to capture even the widest Gaussian beams we test, with a resolution of 3501 by 3501 points, to be well over the Nyquist sampling rate for our wavelength. We take the geometry of Fig.~\ref{fig:schematic} with an ion height of $40$ $\mu $m, though we verify numerically that $\eta$ as calculated vectorially is unaffected by the choice of integration plane, as indicated by equation \ref{eq:field projection}.

We note that the two terms in the overlap integral (Eq. \ref{eq:overlapIntegral}) are equal, so we numerically compute it with the following:
$$ \abs{ \int_{z=z_g} \frac12 \left( \mathbf{E_\mathrm{d}} \times \mathbf{H_\mathrm{g}^*}  \right) \cdot d\mathbf a }^2 
 \label{eq:overlapIntegral 1 term}
$$

For computing the paraxial gaussian, we use $R_{inv} \equiv 1/R$ to bypass the numerical singularity in the intermediate steps.

% \section{Optimal Gaussian giving Clipping}
% \begin{figure}
%     % \centering
%     \includesvg[width= \columnwidth]{Figs/BestWidthPerAperture.svg}
    
%     \caption{Optimal Gaussian given an aperture. Given a circular aperture directly below the ion, the optimal fill factor for the Gaussian beam is to have a radius that is 88\% of the aperture size. Note however at small radii on the grating plane this becomes impossible, because decreasing the beam width further would cause the beam to diverge when exiting the grating.  }
%     \label{fig: optimal gaussian}
% \end{figure}

\end{document}